\documentclass[amsmath,amssymb,preprint,preprintnumbers,floatfix,showpacs]{revtex4}
\usepackage{epsf}
\usepackage{graphicx}
\usepackage{dcolumn}
\usepackage{bm}

\begin{document}

\title{NASCA LINES: A MYSTERY WRAPPED IN AN ENIGMA}
\author{J. R. Castrej\'on-Pita, A. A. Castrej\'on-Pita, A. Sarmiento-Gal\'an$^1$}
\affiliation{Instituto de Matem\'aticas, UNAM, Ave. Universidad s/n,
62200 Chamilpa, Morelos, M\'exico\\
$^1$ e-mail: {\tt ansar@matcuer.unam.mx}}
\author{R. Castrej\'on-Garc{\'\i}a}
\affiliation{Instituto de Investigaciones El\'ectricas, Ave. Reforma 113, 62490
Temixco, Morelos, M\'exico\\}
\vskip0.5in
\date{\today}

\begin{abstract}
We analyze the geometrical structure of the astonishing Nasca geoglyphs in 
terms of their fractal dimension with the idea of dating these manifestations 
of human cultural engagements in relation to one another. Our findings suggest 
that the first delineated images consist of straight, parallel lines and that 
having sophisticated their abilities, Nasca artist moved on to the design of 
more complex structures.
\end{abstract}


\maketitle

{\bf Trying to gain knowledge from the diverse geometrical structure of the 
geoglyphs in the central Peruvian Andes, we have analyzed the biomorphic 
figures in terms of the dimension of each figure. A first finding shows that 
these glyphs have a non integer (fractal) dimension and that its value allows 
a classification of the figures. This result and the reasonable hypothesis 
that the ability of the artists who created them increased with time, leads to 
a chronological ordering of the biomorphs that shows an interesting evolution 
from single straight lines to parallel connected lines and to curvilinear 
nonparallel drawings before designing whole plane trapezoids.}

\section{Introduction}

The well-known Nasca geoglyphs in the central Andes of Per\'u ($14^o 40'$ to 
$14^o 55'$ S lat., $75^o 00'$ to $75^o 10'$ W long.) have been one of the 
XX-th century rediscoveries that remain a mystery in many ways; our study of 
the lines that form these artistic figures shows that they may also be a 
riddle to modern mathematics. These geoglyphs were made by the pre-Inca 
inhabitants of the dry pampa by removing all the dark material (mainly stones) 
and revealing the clear granular substratum. Some indirect methods indicate 
that the geoglyphs were made before the year $600$ (pottery shreds on the 
surfaces~\cite{Silver}), $525 \pm 80$ (radiocarbon of a wooden post found 
nearby~\cite{Reiche}), or from $190$ BC to $660$ (organic material 
``varnish''~\cite{Dorn}); but due to their constitution, it is not possible to 
apply the usual direct dating methods and therefore, one is forced to 
look for alternative ways to estimate their age. The Nasca region has been 
subjected to immense perturbations (ignorant tourism mainly) and some of the 
geoglyphs are terribly damaged; although fortunately there are some good 
photographs that were taken when the figures had just been rediscovered, 
presumably by the mathematician Maria Reiche~\cite{Reiche}, who donated her 
lifelong work to humankind. 

Research in various and diverse fields has shown that there is a tendency for 
some natural systems that evolve with time to increase their 
complexity~\cite{Boyajian}; animal physiology and behavior~\cite{Nat1}, and 
human cultural manifestations not being the exception~\cite{Nat2,Nat3}. Having 
observed that there are different degrees of simplicity in the Nasca figures, 
even for the subset of biomorphs with deca- and hectometric dimensions, we 
have tried to establish their relative ages by measuring their fractal 
dimension. From our present day perspective, we conclude that simpler figures 
would have to be older than more complex ones, in agreement with the 
assumption of a tendency to greater complexity as their creators perfected 
their activity in time.

\section{Method}

Before measuring the fractal dimension of the available figures: three birds, 
two whales, a dog, a monkey, a spider, a pair of hands, and a tree (some shown 
in Fig.~1), we applied a careful digital process that involved 
intensity filtering, contrast magnification, and background-noise elimination, 
that resulted in a significantly increased contour-definition of the 
photographic images; two other methods for processing the figures were also 
employed to confirm consistent results. 

The determination of the fractal dimension for each figure was carried out 
by applying the widely used box-counting method~\cite{Yorke,Nezadal}. Briefly, 
the dimension of a black object over a white background or {\it vice versa}, 
and denoted by $D$, is defined as:
\begin{equation}
D \ = \ \lim_{\varepsilon \to 0} \ 
\frac{\ln N(\varepsilon)}{\ln (1/\varepsilon)},\label{dim} 
\end{equation}
where $N(\varepsilon)$ is the number of boxes in a square grid of side-size 
$\varepsilon $ required to cover the object in question. In our case, the side 
of the box was increased from one image pixel ($0.32 \ mm$) to the whole image 
size ($72 \ dpi$ density resolution), and all images were previously scaled to 
a box $160 \ mm$ wide (the corresponding minimum height is $220 \ mm$); this 
enabled us to avoid the loss of details when the small scales were analyzed. 
Fig.~2 shows the fitting of two straight lines to the measurements 
obtained at different scales. As a uniform rule for avoiding the uncertainties 
due to upper and lower scale cut-offs, we discarded the three lowest and three 
highest points in each graph before performing the fitting; the linear 
correlation coefficients for the ten analyzed figures varied from $0.995$ to 
$0.999$. Our results clearly show discernible differences in the values of 
the fractal dimension of distinct figures that imply an ordering of the 
biomorphs (Fig.~3). We will now use these differences and try to 
establish a chronological ordering of the figures in the following section.

\section{Results and Conclusions}

If one now adopts the point of view~\cite{Tay1,Tay2} that complexity and 
richness increases steadily from dimension one to dimension two, and the 
hypothesis of an increasing complexity in drawing as time evolves is accepted, 
one then has to place lines before trapezoids in time, {\it i. e.}, lines 
older than plane figures (Fig.~3). This result is in agreement with 
some recent archaeological findings~\cite{Aveni,Morrison}, and indicates that 
the Nasca artists started their cultural activity with straight lines and, 
acquiring experience, moved on to the use of more complex geometries. The 
chronological order shown in Fig.~3 places the figures which essentially are 
combinations of straight, parallel lines before the ones that use mainly 
curves and nonparallel draws; the suggested relationship would then imply that 
the first figures are older and less sophisticated than the second set of 
figures, which look more complex in terms of the geometric elements just 
mentioned.

Although our study goes a different way when using the results --trying to 
establish a chronological ordering-- it is similar to the analysis carried out 
for J. Pollock's paintings which enabled the establishing of an initial link 
between the visual preference or aesthetic quality and the value of the 
fractal dimension of an object; it has also opened a new field of research on 
the physiological effects of visual perception (see~\cite{TaylorSA} for a 
review).

We do not expect our premises to be accepted as unquestionable but await for 
them to be contested and challenged; all we can ascertain at present, is that 
the Nasca imagery thus reinflames the debate on the meaning of art, a problem 
which may well continue as an elusive question even for modern mathematical 
tools.

The authors sincerely acknowledge the invaluable suggestions made by R. P. 
Taylor.

\vfill\eject



\begin{figure}[!ht]
\vskip-0.6in
\begin{center}
\epsfxsize=3.5in
\epsffile{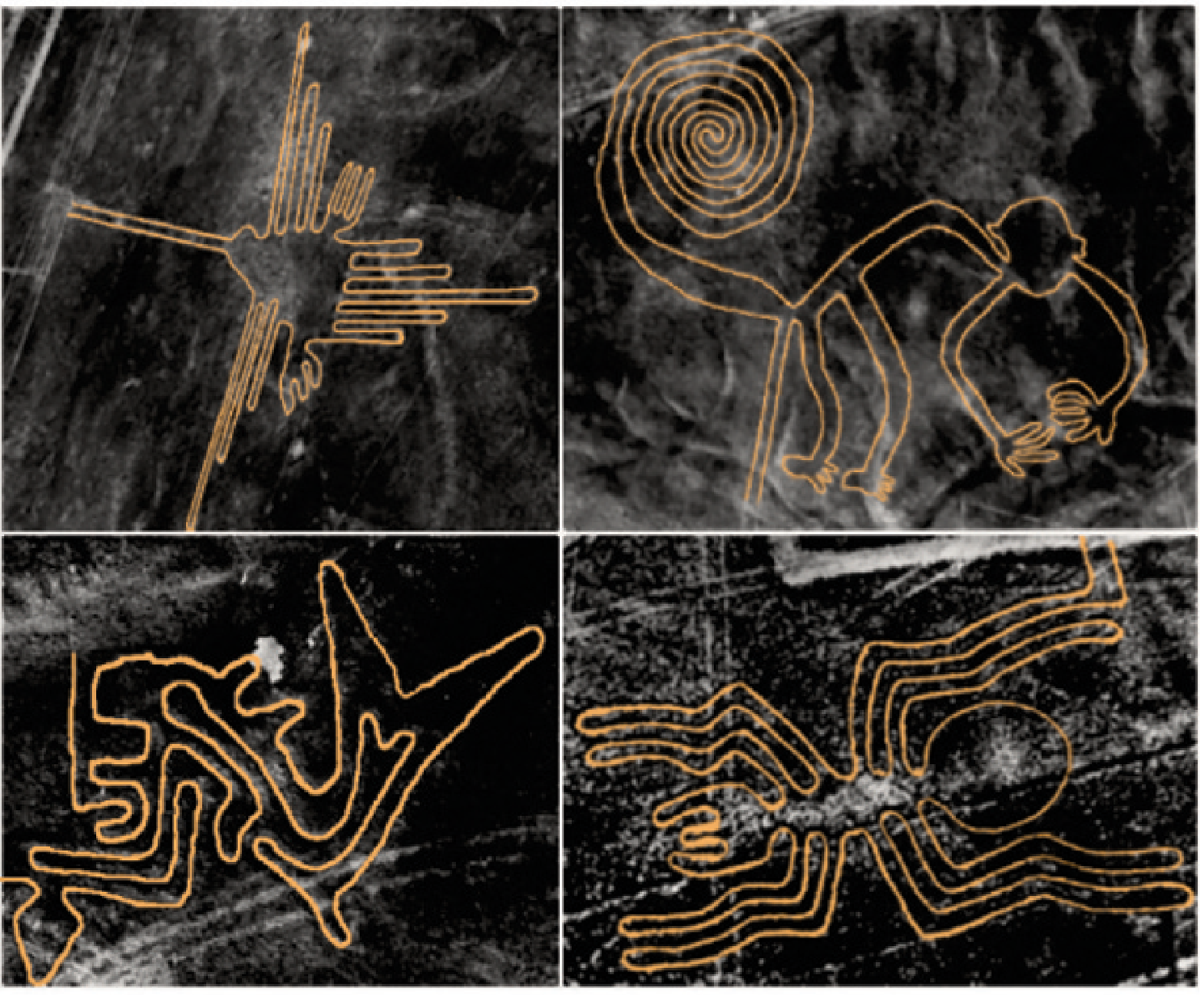}
\end{center}
\vskip-0.9in
\caption{Mosaic showing four of the analyzed biomorphs: humming bird, monkey, 
whale, and spider. The humming bird has a wingspan of $66 \ m$, the actual 
size of the monkey is $135 \ m$ across, while the spider is $46 \ m$ 
long~\cite{Hading}.}
\label{fig1}
\end{figure}
\vfill\eject

\begin{figure}[!ht]
\vskip-0.7in
\begin{center}
\epsfxsize=3.5in
\epsffile{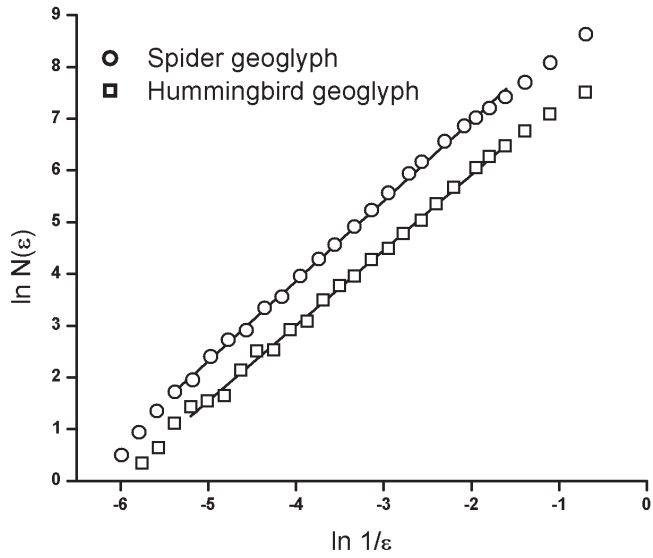}
\end{center}
\caption{Results of the box-counting method for two of the biomorphs in 
Fig.~\ref{fig1}. The extension of the straight lines indicates the data used 
for the linear regression fittings.}
\label{fig2}
\end{figure}
\vfill\eject

\begin{figure}[!ht]
\vskip-0.7in
\begin{center}
\epsfxsize=3.5in
\epsffile{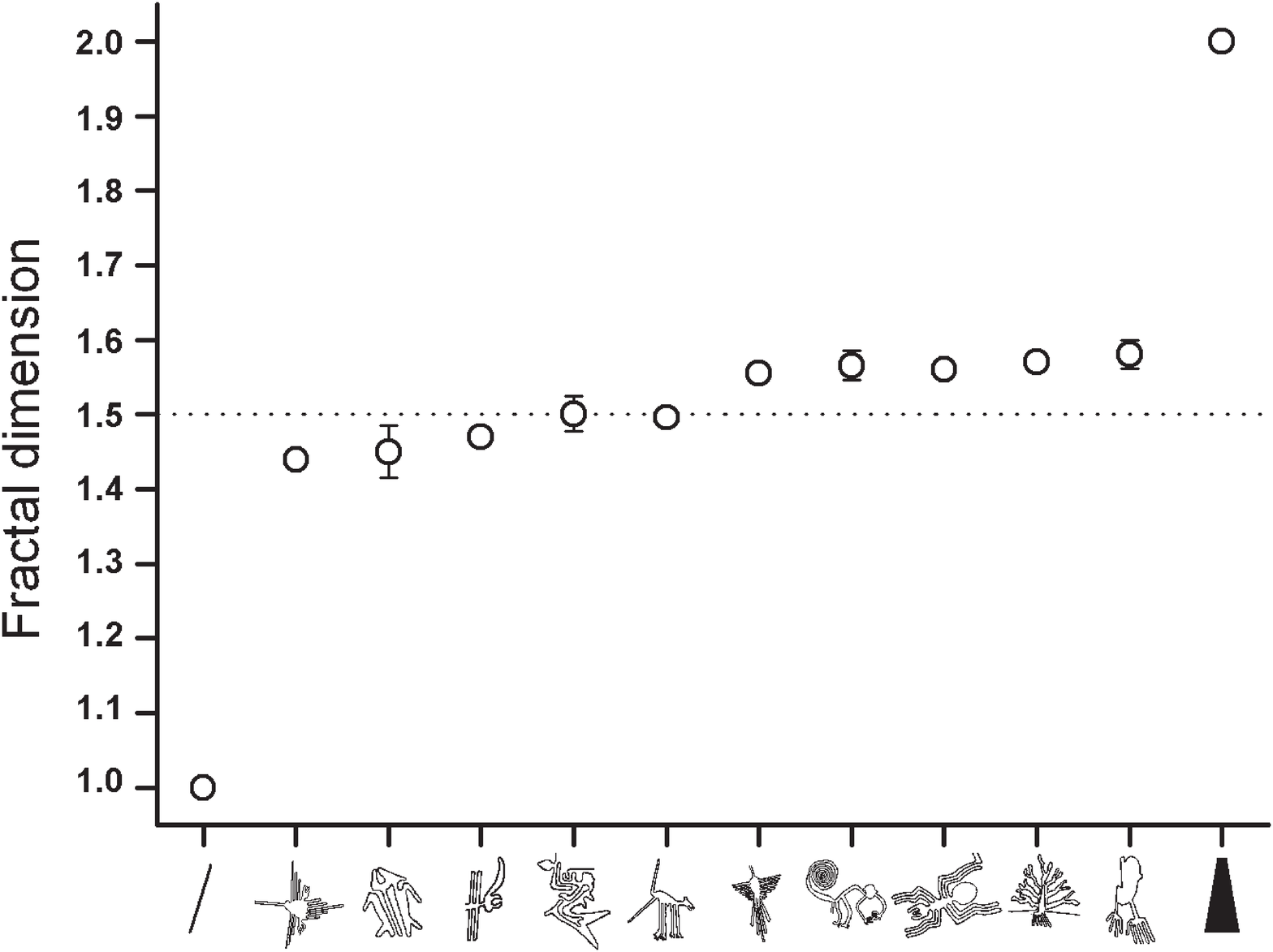}
\end{center}
\caption{Chronological order of the Nasca biomorphs according to their fractal 
dimension.}
\label{fig3}
\end{figure}


\begin{thebibliography}{1}

\bibitem{Silver} H. Silverman, {\it Memoirs of The American Philosophical 
Society}, {\bf 183}, 209 (1990).

\bibitem{Reiche} M. Reiche, {\it Mystery on the Desert}, (Eigenverlag, 
Stuttgart, 1968); {\it Geheimnis der W\"uste}, (Offizin druk: Stuttgart, 1968).

\bibitem{Dorn} R. I. Dorn, {\it American Scientist}, {\bf 79}, 542 (1990).

\bibitem{Boyajian} G. Boyajian, G, T. Lutz, {\it Geology}, {\bf 20}, 983-986 
(1992).

\bibitem{Nat1} J. P. Haskell, M. E. Ritchie, H. Olff, {\it Nature}, {\bf 418}, 
527 (2002).

\bibitem{Nat2} R. P. Taylor, A. P. Micolich, D. Jonas, {\it Nature}, 
{\bf 399}, 422 (1999).

\bibitem{Nat3} M. Kemp, {\it Nature}, {\bf 404}, 546 (2000).

\bibitem{Yorke} K. T. Alligood, T. D. Sauer, J. A. Yorke, {\it Chaos: 
Introduction to Dynamical Systems}, (Springer-Verlag, New York, 1996).

\bibitem{Nezadal} O. Zmeskal, M. Nezadak, and M. Buchnicek, {\it Chaos, 
Solitons and Fractals}, {\bf 17}(1), 113 (2003); Nezadal, M. and Zmeskal O, 
{\it Harmonic and Fractal Image Analyzer}: 
{\tt http://www.fch.vutbr.cz/lectures/imagesci} (2001).

\bibitem{Tay1} R. Taylor, {\it Nature}, {\bf 410}, 18 (2001).

\bibitem{Tay2} R. Taylor, {\it Nature}, {\bf 415}, 961 (2002).

\bibitem{Aveni} A. Aveni, {\it Memoirs of The American Philosophical Society}, 
{\bf 183}, 1 (1990); {\it Between the Lines}, (University of Texas Press, 
Austin, 2000).

\bibitem{Morrison} T. Morrison, {\it Pathways to the Gods}, (Academy, Chicago, 
1988).

\bibitem{TaylorSA} Richard P. Taylor, {\it Scientific American}, {\bf 287}(6), 
116 (Dec., 2002).

\bibitem{Hading} Evan Hadingham, {\it Lines to the Mountain Gods: Nasca and 
the Mysteries of Per\'u}, (University of Oklahoma Press, 1988).

\end{thebibliography}
\end{document}